\documentclass[pra,%
reprint,
 amsmath,amssymb,
 aps,
]{revtex4-2}
\usepackage{amsmath}

\usepackage{graphicx}
\usepackage{dcolumn}
\usepackage{bm}

\begin{document}

\preprint{APS/123-QED}

\title{Cracking Down on Fracture to Functionalise Damage}

\author{Leo~de Waal}
 \email{ldewaal@ed.ac.uk}
\author{Matthaios~Chouzouris}%
\author{Marcelo~A.~Dias}%
\email{mdias@ed.ac.uk}
 \affiliation{Institute for Infrastructure and Environment, School of Engineering, University of Edinburgh}

\date{\today}

\begin{abstract}




In this work we propose a novel relationship between topology and damage propagation in Maxwell lattices that redefines fracture as a functional design feature rather than mere degradation. We demonstrate that topologically protected modes, inherently robust against perturbations, localise along lattice discontinuities and govern the mechanical response. By precisely engineering the microstructure, we direct these modes to control stress distributions and trigger predictable, controlled damage. Our findings---validated through comprehensive numerical simulations and experiments---advance our understanding of nontrivial mechanical responses in Maxwell lattices and establish a clear framework for designing materials with improved fracture energy. This work paves the way for further exploration of topology-driven phenomena in mechanical systems and promises a new direction in the design of robust materials.
\end{abstract}

\maketitle


Designing mechanism of damage in metameterial provides a pathway towards tuneable energy dissipation. Whilst the underlying microstructure of lattice-based metamaterials has been related to the initial fracture toughness~\citep{gibson2003cellular,chen1998fracture,fleck2007damage,tankasala20152013,ATHANASIADIS2021101411,BerkacheKamel2022Meot,omidi2023fracture,choukir2023role}, there is limited understanding of its effect on the subsequent crack growth, during which there is a great latent potential to tune the energy of fracture ~\citep{schmidt2001ductile,tankasala2020crack,hsieh2020versatile,wang2024superior,hedvard2024toughening}. 

Efforts to design the damage path have generally focused on the inclusion of heterogeneities ~\citep{gao2020crack,manno2019engineering,domino2024fracture, gao2024damage,karapiperis2023prediction}. However, recently, Maxwell lattices have shown potential in providing an alternative avenue through their robust topological and geometric quantities~\citep{kane2014topological,zhang2018fracturing,liu2023stress,widstrand2023stress,widstrand2024robustness,dewaal2024architectingmechanismsdamagetopological,Wang2024Topologicalmechanicalmetamaterial}. Specifically, the underlying microstructure can be adjusted to manipulate the form of non-trivial zero-energy deformations \textit{floppy modes} (FM) and \textit{states of self-stress} (SSS) that localise along bulk discontinuities, thus controlling the deformation and stresses that develop within the lattice, particularly at larger \emph{Slenderness Ratio's} (SR - unit-cell size to strut thickness)~\citep{dewaal2024architectingmechanismsdamagetopological}. Furthermore, Maxwell lattices straddle the rigidity transition, providing a pathway to design toughening mechanisms through a change in the failure characteristic~\citep{driscoll2016role}. Such toughening mechanisms are observed in other lattices~\citep{hedvard2024toughening} where zones of bridging elements remain behind the damage path.

\begin{figure}
\includegraphics{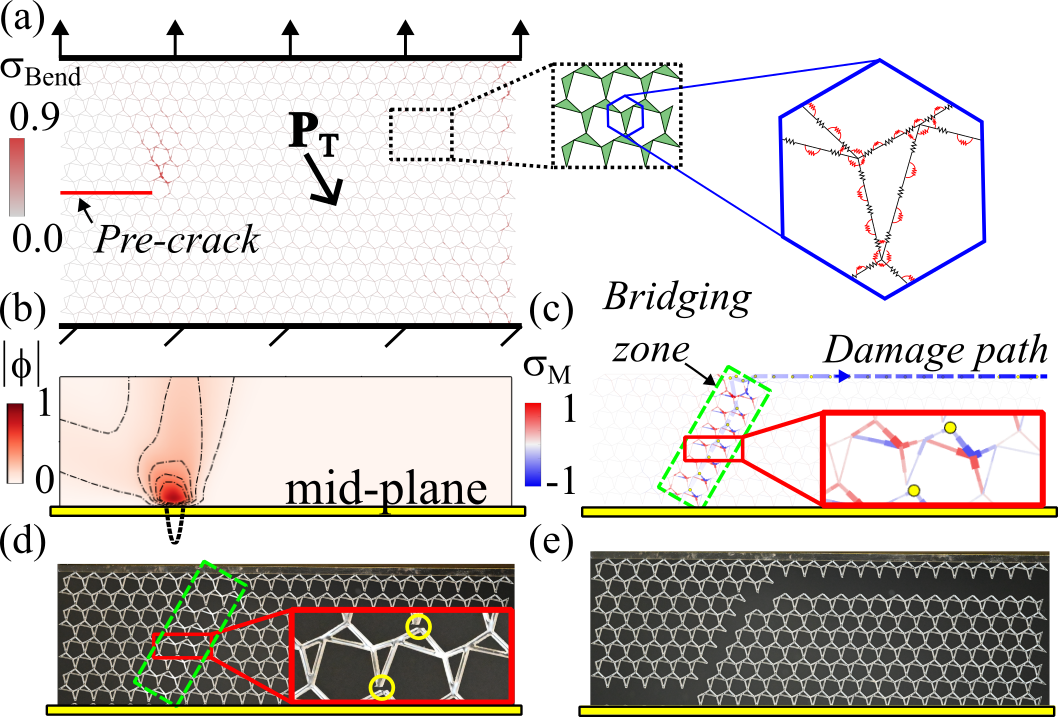}
\caption{\label{fig:1} RDP1 lattice behaviour. (a) Normalised bending stress ($\sigma_\mathrm{Bend} = \left|\sigma_\mathrm{B}/\sigma_\mathrm{B}^\mathrm{u}\right|$) preceding the first failure event with boundary condition and discretisation overview. (b) Mode visualisation for localised disturbance at crack tip with magnitude $\left|\phi\right| = 1$. Details of the mode visualisation are provided in \cite{dewaal2024architectingmechanismsdamagetopological}. (c) Numerical and (d) experimental observation of initial damage path with bridging elements remaining. Zones where bridging elements are present are indicated with a green box. In (c), $\sigma_\mathrm{M} = \sigma_\mathrm{A}/\sigma_\mathrm{A}^\mathrm{u}+\mathrm{sgn}(\sigma_\mathrm{A})\left|\sigma_\mathrm{B}/\sigma_\mathrm{B}^\mathrm{u}\right|$ plots the maximum normalised stress within each element and yellow circles represent locations where the node has split. (e) Complete fracture of sample.}
\end{figure}

In this letter, we explore how the SSS form, damage path, and change in failure characteristics can all be leveraged to increase the fracture energy for pre-cracked distorted kagome lattices. The core idea is presented in Fig.\ref{fig:1} for mode I type tension loading setup. The form of the FM and SSS influences the stress that develops within the lattice. Under fixed boundary conditions, SSS tending towards the long-wavelength limit generally dominate the stresses developing within the lattice, resulting in an overall stretching dominated response~\citep{dewaal2024architectingmechanismsdamagetopological}. However, the bending stresses about the crack tip (Fig.\ref{fig:1}(a)) tend to map onto the FM that localise on the pre-crack edge (Fig.\ref{fig:1}(a) vs (b)). Therefore, when the fracture of the individual elements is dominated by bending, the crack path also aligns with these modes (Fig.\ref{fig:1}(c-e)), providing a methodology to steer the direction of damage. As individual elements fracture, constraints are removed, necessarily resulting in the removal of SSS. However, bridging elements, which deform by bending (Fig.\ref{fig:1}(c,d)), remain along the crack path. These bridging elements provide an opportunity to tune the toughening. Here, we explore this phenomenon numerically and experimentally to elucidate how the form of these topologically protected edge modes can be leveraged to increase fracture energy.

\begin{figure*}
\includegraphics{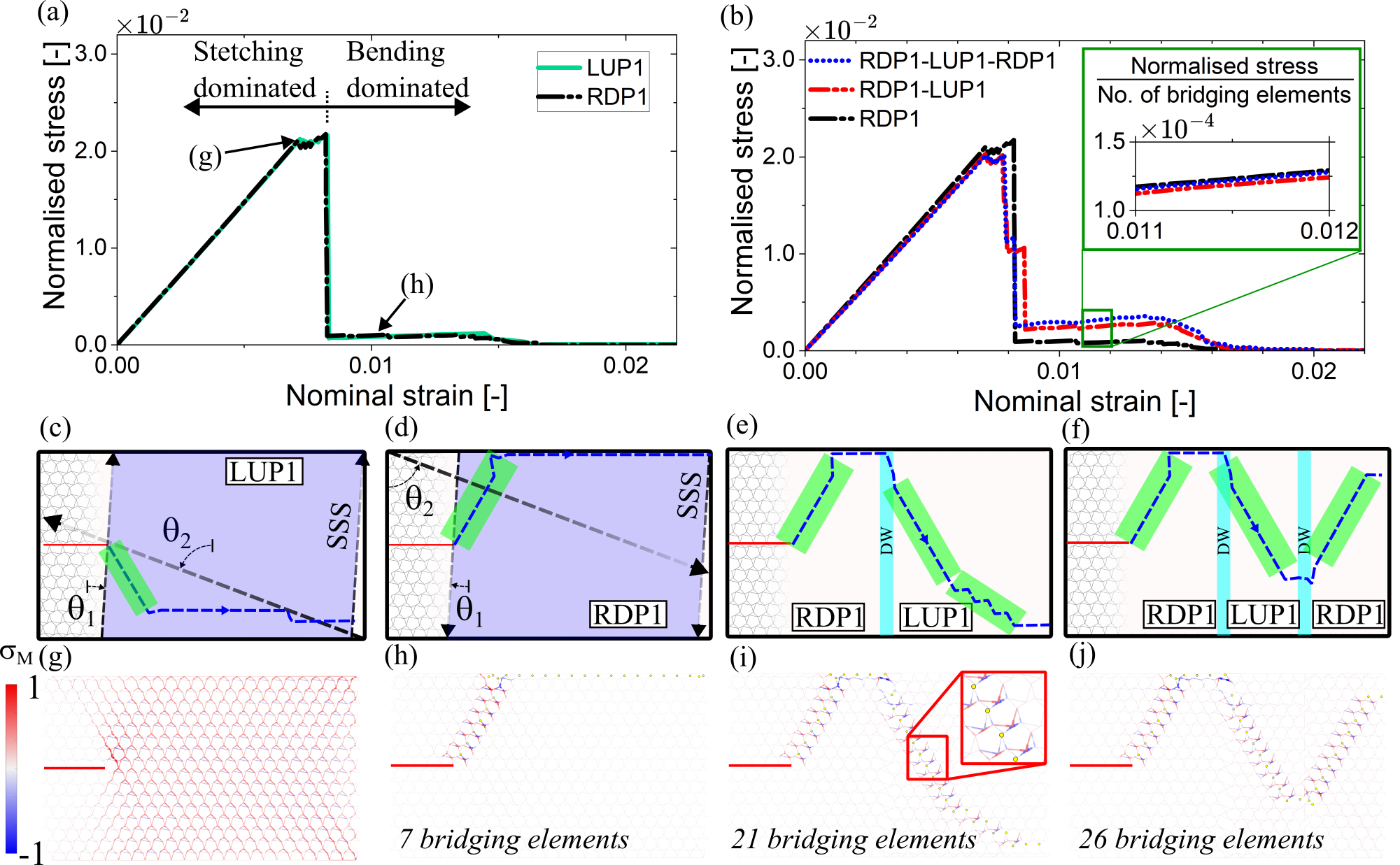}
\caption{\label{fig:2} Normalised stress vs nominal strain for (a) equivalence class LUP1 and RDP1; and, (b) their combination to increase the damage path length. In (b), the insert normalises the stress with the number of bridging elements remaining (provided in (h-j)) in this phase.  (c-f) The damage path with bridging element zones is highlighted in green for LUP1, RDP1, RDP1-LUP1, and RDP1-LUP1-RDP1, respectively. (g) Normalised stress in elements preceding the first failure event for RDP1. Normalised stress following initial stretching dominated damage for (h) RDP1; (i) RDP1-LUP1; and, (j) RDP1-LUP1-RDP1. In (g-j), $\sigma_\mathrm{M} = \sigma_\mathrm{A}/\sigma_\mathrm{A}^\mathrm{u}+\mathrm{sgn}(\sigma_\mathrm{A})\left|\sigma_\mathrm{B}/\sigma_\mathrm{B}^\mathrm{u}\right|$ plots the maximum normalised stress within each element and yellow circles represent locations where the node has split. Refer to Fig. \ref{fig:4} for experimental comparison.}
\end{figure*}

\textit{The role of the microstructure on topologically protected edge modes -} periodic Maxwell lattices are on the verge of mechanical stability, containing an equal number of FM and SSS. In practice, a finite lattice specimen will introduce or remove constraints at its boundary, resulting in an imbalance (loss or gain) between FM and SSS. For example, Maxwell lattices with pinned boundaries (\emph{i.e.}, fixed displacements at the boundary) and a \textit{gapped} dispersion relation will possess extra SSS that exponentially localise around bulk discontinuities~\citep{kane2014topological}. Importantly, these edge modes can exhibit a topologically protected directional preference, known as a topological polarisation $\textbf{P}_\mathrm{T}$ (Fig.\ref{fig:1}(a)), which arises from and depends on the unit-cell's geometry.

We will focus on the response of distorted kagome lattices that can be parameterised in terms of the \emph{regular kagome} (RK) lattice using four parameters $\{x_\mathrm{1}, x_\mathrm{2}, x_\mathrm{3}, z_\mathrm{s}\}$,
\begin{equation}\label{Eq:2}\textbf{r}_\mathrm{\mu} = \textbf{r}_\mathrm{\mu}^\mathrm{0}-\sqrt{3}x_\mathrm{\mu}\textbf{p}_\mathrm{\mu}+x_\mathrm{\mu-1}\textbf{a}_\mathrm{\mu+1}+\frac{z_\mathrm{s}}{\sqrt{3}}\textbf{p}_\mathrm{\mu-1},\end{equation} 
where $\textbf{r}_\mathrm{\mu}^\mathrm{0}$ are the basis vectors for the sites of the RK, $\textbf{p}_\mathrm{\mu}$ are the vectors normal to the bonds, $\textbf{a}_\mathrm{\mu}$ are the primitive lattice vectors, and $\mu\in\{1,2,3\}$~\citep{mao2018maxwell,lubensky2015phonons}. A description of the lattice using the unit-cell defined by Eq.~\eqref{Eq:2} allows the topological polarization to be fully captured by a simple expression only depending on the sign of $x_\mu$~\citep{kane2014topological} when $\mathbf{a}_\mathrm{\mu}$ is fixed:
\begin{equation}\label{Eq:3} \textbf{P}_\mathrm{T} = \frac{1}{2} \sum_\mathrm{\mu=1}^\mathrm{3} \textbf{a}_\mathrm{\mu} \mathrm{sgn}(x_\mathrm{\mu}), \end{equation}
providing a unique definition of topologically distinct lattices through $\mathrm{sgn}(x_\mathrm{1}, x_\mathrm{2}, x_\mathrm{3})$, distinguishing between the different equivalences classes.

While $\textbf{P}_\mathrm{T}$ remains identical within the same equivalence class, the form (\emph{i.e.} direction and decay) of the non-trivial modes can be tuned. Applying the zero-energy condition, expressed in Fourier Space as
\begin{equation} \label{Eq:S5}
    E(q_\mathrm{1}, q_\mathrm{2}) = \det\mathbf{C}(q_\textrm{1},q_\textrm{2}) = \det\mathbf{Q}(q_\textrm{1},q_\textrm{2}) = 0,
\end{equation}
allows the form of these modes in the bulk to be determined in terms of their form on an edge; where $\mathbf{C}$ is the compatibility matrix, $\mathbf{Q}$ is the equilibrium matrix and $q_\mathrm{1,2}$ are wave numbers in the $1$ and $2$-directions. In other words, this energy constraint amounts to fixing one wavenumber component in terms of the other, effectively reducing the relevant degrees of freedom for the description of these evanescent modes - this gives rise to what is known as \textit{bulk-boundary correspondence}. In essence, an evanescent mode that is described by a real wavenumber $q_\mathrm{1}$ parallel to a boundary will be associated with a wavenumber $q_\mathrm{2} \equiv q_\mathrm{2}(q_\mathrm{1})$ orthogonal to it~\citep{kane2014topological}. If the dispersion relation is gapped, the solution for the orthogonal component will be complex, $q_\mathrm{2} = k_\mathrm{2} + i\, \kappa_\mathrm{2}$, where $k_\mathrm{2}$ is the real part of the wavenumber and $\kappa_\mathrm{2}$, its imaginary part, which represents an inverse decay length in the $2$-direction. Therefore, a sinusoidal disturbance decays into the bulk with respect to the edge's inward facing normal at an angle given by \begin{equation}\label{Eq:4} \theta = \arctan\left(\frac{k_\mathrm{2}}{q_\mathrm{1}}\right), \end{equation} 
with clockwise rotation for a negative angle. As $\mathbf{C} = \mathbf{Q}^\dagger$, the $q_\mathrm{2}$ component of the SSS will be the complex conjugate of the FM resulting in $\kappa^\mathrm{SSS} = -\kappa^\mathrm{FM}$ and $\theta^\mathrm{SSS} = \theta^\mathrm{FM} = \theta$. While the sign of $\kappa$ is dependent on $\mathbf{P}_\mathrm{T}$, $|\kappa|$ and $\theta$ are dictated by the underlying cell geometry, allowing lattices even from the same equivalence class to have disparate forms for these modes.

This theory assumes idealised (zero-thickness hinges), however, it has been shown that at larger SR ($>$ 10) the damage path aligns with these modes for non-idealised lattices from equivalence classes (+,-,-) and (-,+,-)~\cite{dewaal2024architectingmechanismsdamagetopological}. We refer to lattices from these equivalence classes as \emph{left-up polarised} (LUP) and \emph{right-down polarised} (RDP) respectively. For both these lattices, $\mathbf{P}_\mathrm{T}$ results in the modes localising exclusively on the bottom/top (Fig.\ref{fig:1}(b)) of the pre-crack. The bending stresses around the crack tip (Fig.\ref{fig:1})a)), which map onto these modes, therefore localise below/above the crack, with the fracture path preferentially following this direction (Fig.\ref{fig:1}(c)). Lattices from the other equivalence classes (\emph{i.e.} ($\pm$,$\mp$,$\mp$) and ($\pm$,$\pm$,$\pm$)) do not show a similar alignment in the fracture path under these loading conditions as both edges of the per-crack harbour a mode. This symmetry results in a horizontal crack path at larger SR. We will, therefore, focus on lattices in the LUP and RDP equivalence classes, as these provide the means to direct the crack path under the selected boundary conditions. Additionally, to ensure that this phenomenon occurs, all lattices studied in this letter will have a constant $\mathrm{SR} = |\mathbf{a}_\mathrm{1}|/t = 15$. Where $t = 0.8mm$ is the bond thickness and $|\mathbf{a}_\mathrm{1}| = 12mm$ is the magnitude of the primitive lattice vector in the $1$-direction. The out-of-plane depth $b = 2.12mm$.  

\textit{Numerical model -} The lattices are simulated using beam elements with fixed boundaries (Fig.\ref{fig:1}(a)). Rotational restraints for this boundary condition are enforced through additional torsional springs attached to the support. The fracture process is modelled by splitting a node at the end of the element(s) when the local failure condition,
\begin{equation}\label{Eq:7} \left|\frac{\sigma_\mathrm{A}}{\sigma_\mathrm{A}^\mathrm{u}}\right|+\left|\frac{\sigma_\mathrm{B}}{\sigma_\mathrm{B}^\mathrm{u}}\right| \geq 1, \end{equation} 
is reached. Where $\sigma_\mathrm{A}$ and $\sigma_\mathrm{B}$ are the maximum axial and bending stresses in the element, and $\sigma_\mathrm{A}^\mathrm{u}$ and $\sigma_\mathrm{B}^\mathrm{u}$ are the material failure stress values in axial and bending respectively. The material is modelled as linear-elastic until failure with elastic modulus, $\sigma_\mathrm{A}^\mathrm{u}$ and $\sigma_\mathrm{B}^\mathrm{u}$ set to 3120MPa, 45MPa and 85MPa. The elastic modulus is taken from the manufacturer's data sheet, whilst the other properties are found from material tests on 2.12mm thick PMMA. Further details of the model can be found in \cite{dewaal2024architectingmechanismsdamagetopological}. When reporting results, the stress applied to the edge of the lattice is normalised with $\sigma_\mathrm{A}^\mathrm{u}$ (\emph{i.e.} $\sigma_\mathrm{n} = F/(A_{\mbox{\tiny edge}} \sigma_\mathrm{A}^\mathrm{u})$ where $F$ is the force applied to the sites and $A_{\mbox{\tiny edge}}$ is the projected area of the edge). The nominal strain is $\epsilon_\mathrm{n} = \delta/H$, where $H$ is the initial height of the lattice.

\begin{figure}
\includegraphics{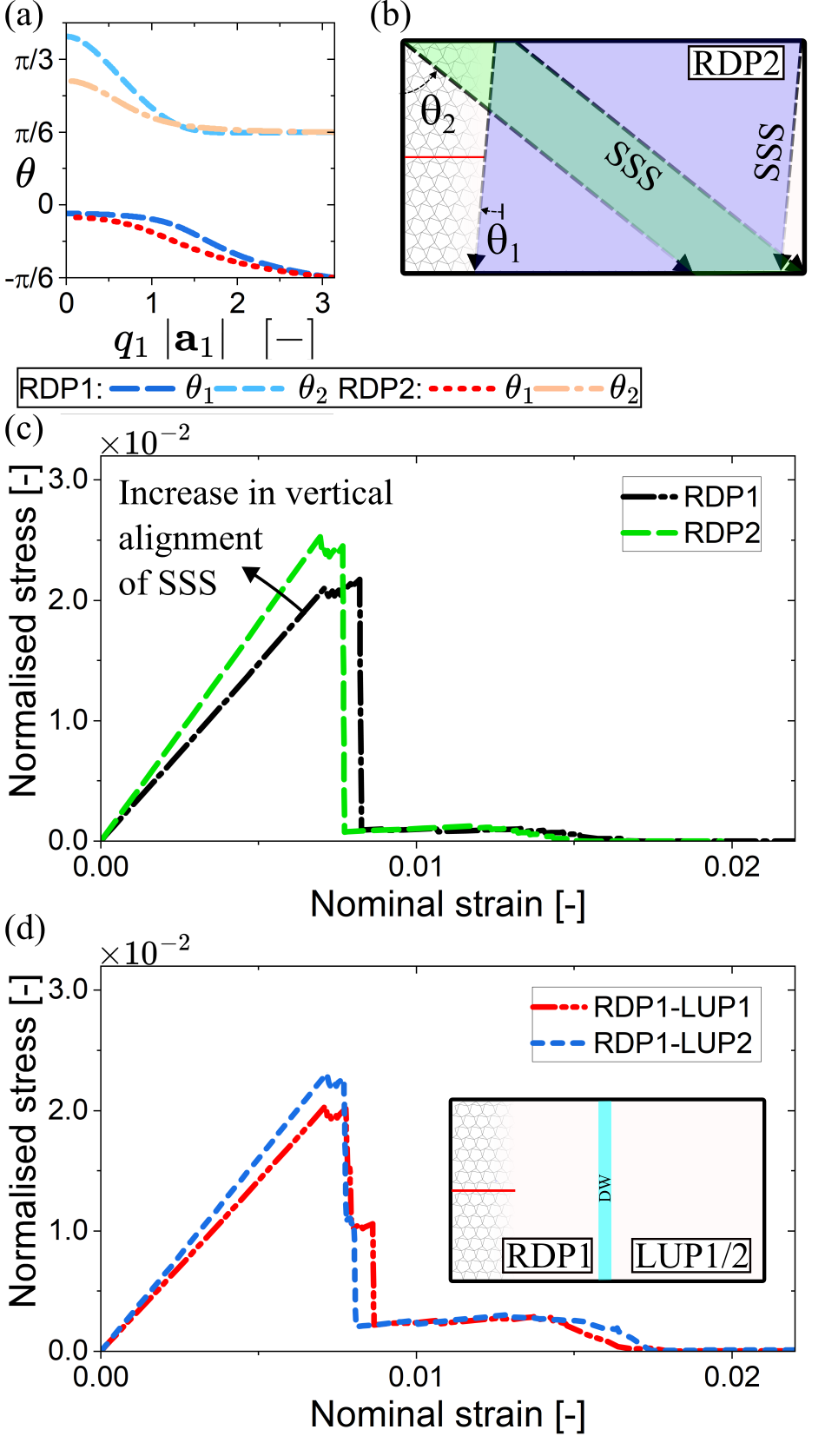}
\caption{\label{fig:3} (a) Direction of modes for RDP1-2. (b) SSS at the long wavelength limit for RDP2. Effect of SSS alignment on normalised stress vs nominal strain without domain walls for (d) RDP1-2 and (e) with domain walls RDP1-LUP1/2.}
\end{figure}

\textit{Numerical results -} RDP1 and LUP1 unit cells are rotationally symmetric with $(|x_\mathrm{1}|, |x_\mathrm{2}|, |x_\mathrm{3}|) = (0.08, 0.08, 0.1), z = 0.$ Both lattices have a similar response (Fig. \ref{fig:2}(a)) and damage process. The damage path initially propagates diagonally downwards/upwards before continuing along a horizontal path upon nearing the boundary, shown with a blue dashed line in Fig. \ref{fig:2}(c,d)).  Around the regions of diagonal damage propagation, bridging elements remain (Fig. \ref{fig:2}(h) and Fig. \ref{fig:1}(d)), highlighted in green zones in Fig. \ref{fig:2}(c,d). Along the horizontal damage propagation, the two surfaces are completely disconnected. The increase in degrees of freedom, related to the initial damage propagation, results in the dominant behaviour switching from stretching to bending. These two regions are indicated in Fig. \ref{fig:2}(a). In the initial stretching dominated phase, the stress in the lattice (Fig. \ref{fig:2}(g)) maps onto the SSS form (Fig. \ref{fig:2}(g)). As the initial damage propagates, these SSS are removed until only the bridging elements, dominated by bending, remain (Fig. \ref{fig:2}(h)). In both phases there is an opportunity to increase the fracture energy by adjusting the microstructure.

Increasing the number of bridging elements provides a methodology to increase the energy absorbed within the second, bending dominated phase. The number of bridging elements is proportional to the length of the diagonal damage path. To increase the diagonal path length, we can combine RDP1 and LUP1 unit-cells, separated by domain walls, to direct the crack away from the boundary in a zig-zag type propagation (Fig. \ref{fig:2}(e,f,i,j)). The initial stretching dominated phase remains similar when adding domain walls, provided they are not positioned to significantly alter the stress profile about the pre-crack, resulting in a similar peak stress and stiffening response (Fig. \ref{fig:2}(b)). However, the stiffness increases linearly with the number of bridging elements remaining within the second phase. The small insert within Fig. \ref{fig:2}(b) normalises the stress-strain response with respect to the number of bridging elements remaining within the second phase, highlighting this scaling.    
%
%

Within the first phase, alignment of the SSS provides a methodology to increase the fracture energy. At the long-wavelength limit, only SSS for $\theta_{1}$ can develop for the domain size simulated (Fig. \ref{fig:2}(d)). RDP2 (0.12,-0.04,0.08) microstructure is adjusted so $\theta_{1}$ remains similar but $\theta_{2}$ is reduced (Fig. \ref{fig:3}(a)), allowing for the SSS associated with this mode to develop within the lattice (Fig. \ref{fig:3}(b)). This increases the initial stiffness (Fig. \ref{fig:3}(c)). However, the initial fracture occurs at a lower nominal strain, reducing the possible increase in fracture energy. By leveraging domain walls, it is possible to increase the stiffness whilst maintaining a similar nominal strain for fracture initiation. This is shown in Fig. \ref{fig:3}(d), which compares the response for two lattices with a single domain walls separating RDP1 and LUP1/2 microstructure. Due to the more vertical alignment of LUP2 when compared to LUP1, the initial stiffness is larger for RDP1-LUP2, and, as RDP1 is located at the pre-crack in both cases, the nominal strain where fracture initiates is identical. The microstructure can be adjusted to reduce/increase the vertical alignment of $\theta_{1}$ for a similar effect. 

\begin{figure}
\includegraphics{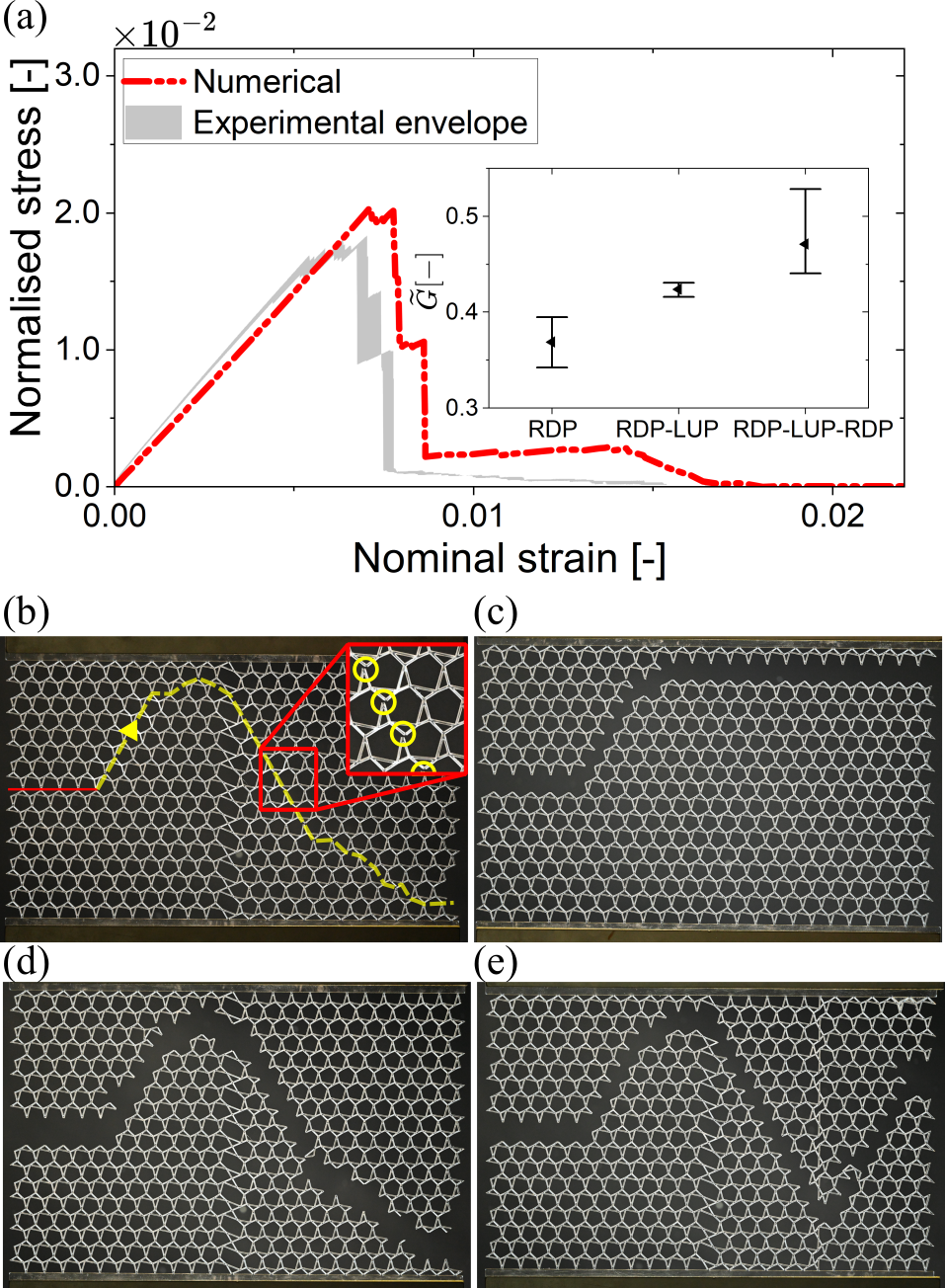}
\caption{\label{fig:4} Numerical vs experimental. (a) Normalised stress vs nominal strain with insert showing the normalised fracture energy $\tilde{G} = G/(\sigma_{\mathrm{A}}^{\mathrm{u}}\left|a_{1}\right|bt)$. $G$ is the energy required to separate the lattice. (b) Initial damage path with bridging elements intact for RDP1-LUP1; and, complete fracture for (c) RDP1, (d) RDP1-LUP1, (e) RDP1-LUP1-RDP1. The numerical prediction of the damage path is provided in Fig. \ref{fig:2}(d-f,h-j).}
\end{figure}

\textit{Experimental observation-} experiments were conducted on samples laser cut from 2.12mm thick PMMA. These samples were annealed for 2hrs at 85$^{\circ}$C to reduce residual stresses. Three samples were tested for each lattice under type I loading with fixed boundaries. A comparison with the simulation for RDP1-LUP1 is provided in Fig. \ref{fig:4}. See Supplementary Material Fig. S1 for the comparison of other lattices. Whilst the numerical model sligtly over-predicts the peak stress, the general behaviour is in close agreement, with the two distinct phases being captured (Fig. \ref{fig:4}(a)). In the experiments, bridging elements are observed following the initial damage propagation (Fig. \ref{fig:4}(b)). These sequentially fracture, leading to the complete separation of the sample (Fig. \ref{fig:4}(d)). In all cases, the damage process predicted numerically was observed experimentally (Fig. \ref{fig:4}(c,e)). The good agreement with experiments suggests that this phenomenology is robust against manufacturing imperfections (this is also supported by a sensitivity study presented in the Supplementary Material). However, whilst the initial magnitude of the stress within the bending dominated phase is similar to that predicted numerically, it reduces below the prediction as loading continues (Fig. \ref{fig:4}(a)). Notwithstanding, the experiments indicate that increasing the damage path results in an increase in the fracture energy (Fig. \ref{fig:4}(a)).  

\textit{Conclusions - } 
This work establishes a robust strategy for functionalising damage in Maxwell lattices by leveraging their topologically protected edge modes. By tuning the form of these modes through the microstructure, we demonstrate how the stress distribution and subsequent damage propagation can be controlled at large SR. We identify two distinct failure stages and introduce a methodology to tailor the fracture energy in each: in the first, stretching-dominated stage, SSS are progressively removed as damage advances, with the vertical alignment of these SSS modulating the lattice’s effective stiffness and offering a key opportunity to enhance fracture energy; in the subsequent, bending-dominated stage, bridging elements persist along the damage path, where the absorbed energy scales with the fracture length. Combining lattice geometries to extend this path thus provides a reliable means to increase energy absorption during failure. The convergence of numerical simulations and experimental validation reinforces the robustness of our approach. These insights not only deepen our understanding of topological modes in governing mechanical responses in Maxwell lattices, but also pave the way for the engineered design of materials with tailored and improved fracture energy. Our work highlights the fundamental role of topology in dictating damage propagation and opens new directions for applying these principles across a broader class of metamaterials and complex structures.

\begin{acknowledgments}
We are grateful to A. Lingua, A. Sanner and D. Kammer from ETHZ for many informative discussions. The authors thank UKRI for support under the EPSRC Open Fellowship scheme (Project No. EP/W019450/1).
\paragraph*{Author contributions:}
L.d.W. and M.A.D. designed the research; M.A.D. supervised and acquired funding; L.d.W. developed the numerical analysis and conducted the experimental investigation; L.d.W., M.C., and M.A.D. performed research; L.d.W., M.C., and M.A.D. analysed data; L.d.W. wrote original draft; and L.d.W., M.C., and M.A.D. wrote and edited the paper.
\end{acknowledgments}



\begin{thebibliography}{28}%
\makeatletter
\providecommand \@ifxundefined [1]{%
 \@ifx{#1\undefined}
}%
\providecommand \@ifnum [1]{%
 \ifnum #1\expandafter \@firstoftwo
 \else \expandafter \@secondoftwo
 \fi
}%
\providecommand \@ifx [1]{%
 \ifx #1\expandafter \@firstoftwo
 \else \expandafter \@secondoftwo
 \fi
}%
\providecommand \natexlab [1]{#1}%
\providecommand \enquote  [1]{``#1''}%
\providecommand \bibnamefont  [1]{#1}%
\providecommand \bibfnamefont [1]{#1}%
\providecommand \citenamefont [1]{#1}%
\providecommand \href@noop [0]{\@secondoftwo}%
\providecommand \href [0]{\begingroup \@sanitize@url \@href}%
\providecommand \@href[1]{\@@startlink{#1}\@@href}%
\providecommand \@@href[1]{\endgroup#1\@@endlink}%
\providecommand \@sanitize@url [0]{\catcode `\\12\catcode `\$12\catcode
  `\&12\catcode `\#12\catcode `\^12\catcode `\_12\catcode `\%12\relax}%
\providecommand \@@startlink[1]{}%
\providecommand \@@endlink[0]{}%
\providecommand \url  [0]{\begingroup\@sanitize@url \@url }%
\providecommand \@url [1]{\endgroup\@href {#1}{\urlprefix }}%
\providecommand \urlprefix  [0]{URL }%
\providecommand \Eprint [0]{\href }%
\providecommand \doibase [0]{https://doi.org/}%
\providecommand \selectlanguage [0]{\@gobble}%
\providecommand \bibinfo  [0]{\@secondoftwo}%
\providecommand \bibfield  [0]{\@secondoftwo}%
\providecommand \translation [1]{[#1]}%
\providecommand \BibitemOpen [0]{}%
\providecommand \bibitemStop [0]{}%
\providecommand \bibitemNoStop [0]{.\EOS\space}%
\providecommand \EOS [0]{\spacefactor3000\relax}%
\providecommand \BibitemShut  [1]{\csname bibitem#1\endcsname}%
\let\auto@bib@innerbib\@empty
\bibitem [{\citenamefont {Gibson}(2003)}]{gibson2003cellular}%
  \BibitemOpen
  \bibfield  {author} {\bibinfo {author} {\bibfnamefont {L.~J.}\ \bibnamefont
  {Gibson}},\ }\bibfield  {title} {\bibinfo {title} {Cellular solids},\
  }\href@noop {} {\bibfield  {journal} {\bibinfo  {journal} {Mrs Bulletin}\
  }\textbf {\bibinfo {volume} {28}},\ \bibinfo {pages} {270} (\bibinfo {year}
  {2003})}\BibitemShut {NoStop}%
\bibitem [{\citenamefont {Chen}\ \emph {et~al.}(1998)\citenamefont {Chen},
  \citenamefont {Huang},\ and\ \citenamefont {Ortiz}}]{chen1998fracture}%
  \BibitemOpen
  \bibfield  {author} {\bibinfo {author} {\bibfnamefont {J.}~\bibnamefont
  {Chen}}, \bibinfo {author} {\bibfnamefont {Y.}~\bibnamefont {Huang}},\ and\
  \bibinfo {author} {\bibfnamefont {M.}~\bibnamefont {Ortiz}},\ }\bibfield
  {title} {\bibinfo {title} {Fracture analysis of cellular materials: a strain
  gradient model},\ }\href@noop {} {\bibfield  {journal} {\bibinfo  {journal}
  {Journal of the Mechanics and Physics of Solids}\ }\textbf {\bibinfo {volume}
  {46}},\ \bibinfo {pages} {789} (\bibinfo {year} {1998})}\BibitemShut
  {NoStop}%
\bibitem [{\citenamefont {Fleck}\ and\ \citenamefont
  {Qiu}(2007)}]{fleck2007damage}%
  \BibitemOpen
  \bibfield  {author} {\bibinfo {author} {\bibfnamefont {N.~A.}\ \bibnamefont
  {Fleck}}\ and\ \bibinfo {author} {\bibfnamefont {X.}~\bibnamefont {Qiu}},\
  }\bibfield  {title} {\bibinfo {title} {The damage tolerance of
  elastic--brittle, two-dimensional isotropic lattices},\ }\href@noop {}
  {\bibfield  {journal} {\bibinfo  {journal} {Journal of the Mechanics and
  Physics of Solids}\ }\textbf {\bibinfo {volume} {55}},\ \bibinfo {pages}
  {562} (\bibinfo {year} {2007})}\BibitemShut {NoStop}%
\bibitem [{\citenamefont {Tankasala}\ \emph {et~al.}(2015)\citenamefont
  {Tankasala}, \citenamefont {Deshpande},\ and\ \citenamefont
  {Fleck}}]{tankasala20152013}%
  \BibitemOpen
  \bibfield  {author} {\bibinfo {author} {\bibfnamefont {H.~C.}\ \bibnamefont
  {Tankasala}}, \bibinfo {author} {\bibfnamefont {V.~S.}\ \bibnamefont
  {Deshpande}},\ and\ \bibinfo {author} {\bibfnamefont {N.~A.}\ \bibnamefont
  {Fleck}},\ }\bibfield  {title} {\bibinfo {title} {2013 koiter medal paper:
  crack-tip fields and toughness of two-dimensional elastoplastic lattices},\
  }\href@noop {} {\bibfield  {journal} {\bibinfo  {journal} {Journal of Applied
  Mechanics}\ }\textbf {\bibinfo {volume} {82}},\ \bibinfo {pages} {091004}
  (\bibinfo {year} {2015})}\BibitemShut {NoStop}%
\bibitem [{\citenamefont {Athanasiadis}\ \emph {et~al.}(2021)\citenamefont
  {Athanasiadis}, \citenamefont {Dias},\ and\ \citenamefont
  {Budzik}}]{ATHANASIADIS2021101411}%
  \BibitemOpen
  \bibfield  {author} {\bibinfo {author} {\bibfnamefont {A.~E.}\ \bibnamefont
  {Athanasiadis}}, \bibinfo {author} {\bibfnamefont {M.~A.}\ \bibnamefont
  {Dias}},\ and\ \bibinfo {author} {\bibfnamefont {M.~K.}\ \bibnamefont
  {Budzik}},\ }\bibfield  {title} {\bibinfo {title} {Can confined mechanical
  metamaterials replace adhesives?},\ }\href@noop {} {\bibfield  {journal}
  {\bibinfo  {journal} {Extreme Mechanics Letters}\ }\textbf {\bibinfo {volume}
  {48}},\ \bibinfo {pages} {101411} (\bibinfo {year} {2021})}\BibitemShut
  {NoStop}%
\bibitem [{\citenamefont {Berkache}\ \emph {et~al.}(2022)\citenamefont
  {Berkache}, \citenamefont {Phani},\ and\ \citenamefont
  {Ganghoffer}}]{BerkacheKamel2022Meot}%
  \BibitemOpen
  \bibfield  {author} {\bibinfo {author} {\bibfnamefont {K.}~\bibnamefont
  {Berkache}}, \bibinfo {author} {\bibfnamefont {S.}~\bibnamefont {Phani}},\
  and\ \bibinfo {author} {\bibfnamefont {J.-F.}\ \bibnamefont {Ganghoffer}},\
  }\bibfield  {title} {\bibinfo {title} {Micropolar effects on the effective
  elastic properties and elastic fracture toughness of planar lattices},\
  }\href@noop {} {\bibfield  {journal} {\bibinfo  {journal} {European journal
  of mechanics, A, Solids}\ }\textbf {\bibinfo {volume} {93}},\ \bibinfo
  {pages} {104489} (\bibinfo {year} {2022})}\BibitemShut {NoStop}%
\bibitem [{\citenamefont {Omidi}\ and\ \citenamefont
  {St-Pierre}(2023)}]{omidi2023fracture}%
  \BibitemOpen
  \bibfield  {author} {\bibinfo {author} {\bibfnamefont {M.}~\bibnamefont
  {Omidi}}\ and\ \bibinfo {author} {\bibfnamefont {L.}~\bibnamefont
  {St-Pierre}},\ }\bibfield  {title} {\bibinfo {title} {Fracture toughness of
  semi-regular lattices},\ }\href@noop {} {\bibfield  {journal} {\bibinfo
  {journal} {International Journal of Solids and Structures}\ }\textbf
  {\bibinfo {volume} {270}},\ \bibinfo {pages} {112233} (\bibinfo {year}
  {2023})}\BibitemShut {NoStop}%
\bibitem [{\citenamefont {Choukir}\ and\ \citenamefont
  {Singh}(2023)}]{choukir2023role}%
  \BibitemOpen
  \bibfield  {author} {\bibinfo {author} {\bibfnamefont {S.}~\bibnamefont
  {Choukir}}\ and\ \bibinfo {author} {\bibfnamefont {C.}~\bibnamefont
  {Singh}},\ }\bibfield  {title} {\bibinfo {title} {Role of topology in
  dictating the fracture toughness of mechanical metamaterials},\ }\href@noop
  {} {\bibfield  {journal} {\bibinfo  {journal} {International Journal of
  Mechanical Sciences}\ }\textbf {\bibinfo {volume} {241}},\ \bibinfo {pages}
  {107945} (\bibinfo {year} {2023})}\BibitemShut {NoStop}%
\bibitem [{\citenamefont {Schmidt}\ and\ \citenamefont
  {Fleck}(2001)}]{schmidt2001ductile}%
  \BibitemOpen
  \bibfield  {author} {\bibinfo {author} {\bibfnamefont {I.}~\bibnamefont
  {Schmidt}}\ and\ \bibinfo {author} {\bibfnamefont {N.}~\bibnamefont
  {Fleck}},\ }\bibfield  {title} {\bibinfo {title} {Ductile fracture of
  two-dimensional cellular structures--dedicated to prof. dr.-ing. d. gross on
  the occasion of his 60th birthday},\ }\href@noop {} {\bibfield  {journal}
  {\bibinfo  {journal} {International Journal of Fracture}\ }\textbf {\bibinfo
  {volume} {111}},\ \bibinfo {pages} {327} (\bibinfo {year}
  {2001})}\BibitemShut {NoStop}%
\bibitem [{\citenamefont {Tankasala}\ and\ \citenamefont
  {Fleck}(2020)}]{tankasala2020crack}%
  \BibitemOpen
  \bibfield  {author} {\bibinfo {author} {\bibfnamefont {H.~C.}\ \bibnamefont
  {Tankasala}}\ and\ \bibinfo {author} {\bibfnamefont {N.~A.}\ \bibnamefont
  {Fleck}},\ }\bibfield  {title} {\bibinfo {title} {The crack growth resistance
  of an elastoplastic lattice},\ }\href@noop {} {\bibfield  {journal} {\bibinfo
   {journal} {International Journal of Solids and Structures}\ }\textbf
  {\bibinfo {volume} {188}},\ \bibinfo {pages} {233} (\bibinfo {year}
  {2020})}\BibitemShut {NoStop}%
\bibitem [{\citenamefont {Hsieh}\ \emph {et~al.}(2020)\citenamefont {Hsieh},
  \citenamefont {Deshpande},\ and\ \citenamefont
  {Valdevit}}]{hsieh2020versatile}%
  \BibitemOpen
  \bibfield  {author} {\bibinfo {author} {\bibfnamefont {M.-T.}\ \bibnamefont
  {Hsieh}}, \bibinfo {author} {\bibfnamefont {V.~S.}\ \bibnamefont
  {Deshpande}},\ and\ \bibinfo {author} {\bibfnamefont {L.}~\bibnamefont
  {Valdevit}},\ }\bibfield  {title} {\bibinfo {title} {A versatile numerical
  approach for calculating the fracture toughness and r-curves of cellular
  materials},\ }\href@noop {} {\bibfield  {journal} {\bibinfo  {journal}
  {Journal of the Mechanics and Physics of Solids}\ }\textbf {\bibinfo {volume}
  {138}},\ \bibinfo {pages} {103925} (\bibinfo {year} {2020})}\BibitemShut
  {NoStop}%
\bibitem [{\citenamefont {Wang}\ \emph
  {et~al.}(2024{\natexlab{a}})\citenamefont {Wang}, \citenamefont {Wu},
  \citenamefont {Zhang}, \citenamefont {Li}, \citenamefont {Wang},\ and\
  \citenamefont {Gao}}]{wang2024superior}%
  \BibitemOpen
  \bibfield  {author} {\bibinfo {author} {\bibfnamefont {Y.}~\bibnamefont
  {Wang}}, \bibinfo {author} {\bibfnamefont {K.}~\bibnamefont {Wu}}, \bibinfo
  {author} {\bibfnamefont {X.}~\bibnamefont {Zhang}}, \bibinfo {author}
  {\bibfnamefont {X.}~\bibnamefont {Li}}, \bibinfo {author} {\bibfnamefont
  {Y.}~\bibnamefont {Wang}},\ and\ \bibinfo {author} {\bibfnamefont
  {H.}~\bibnamefont {Gao}},\ }\bibfield  {title} {\bibinfo {title} {Superior
  fracture resistance and topology-induced intrinsic toughening mechanism in 3d
  shell-based lattice metamaterials},\ }\href@noop {} {\bibfield  {journal}
  {\bibinfo  {journal} {Science Advances}\ }\textbf {\bibinfo {volume} {10}},\
  \bibinfo {pages} {eadq2664} (\bibinfo {year}
  {2024}{\natexlab{a}})}\BibitemShut {NoStop}%
\bibitem [{\citenamefont {Hedvard}\ \emph {et~al.}(2024)\citenamefont
  {Hedvard}, \citenamefont {Dias},\ and\ \citenamefont
  {Budzik}}]{hedvard2024toughening}%
  \BibitemOpen
  \bibfield  {author} {\bibinfo {author} {\bibfnamefont {M.~L.}\ \bibnamefont
  {Hedvard}}, \bibinfo {author} {\bibfnamefont {M.~A.}\ \bibnamefont {Dias}},\
  and\ \bibinfo {author} {\bibfnamefont {M.~K.}\ \bibnamefont {Budzik}},\
  }\bibfield  {title} {\bibinfo {title} {Toughening mechanisms and damage
  propagation in architected-interfaces},\ }\href@noop {} {\bibfield  {journal}
  {\bibinfo  {journal} {International Journal of Solids and Structures}\
  }\textbf {\bibinfo {volume} {288}},\ \bibinfo {pages} {112600} (\bibinfo
  {year} {2024})}\BibitemShut {NoStop}%
\bibitem [{\citenamefont {Gao}\ \emph {et~al.}(2020)\citenamefont {Gao},
  \citenamefont {Li}, \citenamefont {Dong},\ and\ \citenamefont
  {Zhao}}]{gao2020crack}%
  \BibitemOpen
  \bibfield  {author} {\bibinfo {author} {\bibfnamefont {Z.}~\bibnamefont
  {Gao}}, \bibinfo {author} {\bibfnamefont {D.}~\bibnamefont {Li}}, \bibinfo
  {author} {\bibfnamefont {G.}~\bibnamefont {Dong}},\ and\ \bibinfo {author}
  {\bibfnamefont {Y.~F.}\ \bibnamefont {Zhao}},\ }\bibfield  {title} {\bibinfo
  {title} {Crack path-engineered 2d octet-truss lattice with bio-inspired crack
  deflection},\ }\href@noop {} {\bibfield  {journal} {\bibinfo  {journal}
  {Additive Manufacturing}\ }\textbf {\bibinfo {volume} {36}},\ \bibinfo
  {pages} {101539} (\bibinfo {year} {2020})}\BibitemShut {NoStop}%
\bibitem [{\citenamefont {Manno}\ \emph {et~al.}(2019)\citenamefont {Manno},
  \citenamefont {Gao},\ and\ \citenamefont {Benedetti}}]{manno2019engineering}%
  \BibitemOpen
  \bibfield  {author} {\bibinfo {author} {\bibfnamefont {R.}~\bibnamefont
  {Manno}}, \bibinfo {author} {\bibfnamefont {W.}~\bibnamefont {Gao}},\ and\
  \bibinfo {author} {\bibfnamefont {I.}~\bibnamefont {Benedetti}},\ }\bibfield
  {title} {\bibinfo {title} {Engineering the crack path in lattice cellular
  materials through bio-inspired micro-structural alterations},\ }\href@noop {}
  {\bibfield  {journal} {\bibinfo  {journal} {Extreme Mechanics Letters}\
  }\textbf {\bibinfo {volume} {26}},\ \bibinfo {pages} {8} (\bibinfo {year}
  {2019})}\BibitemShut {NoStop}%
\bibitem [{\citenamefont {Domino}\ \emph {et~al.}(2024)\citenamefont {Domino},
  \citenamefont {d'Aug{\`e}res}, \citenamefont {Zhang}, \citenamefont {Janbaz},
  \citenamefont {Arag{\`o}n},\ and\ \citenamefont
  {Coulais}}]{domino2024fracture}%
  \BibitemOpen
  \bibfield  {author} {\bibinfo {author} {\bibfnamefont {L.}~\bibnamefont
  {Domino}}, \bibinfo {author} {\bibfnamefont {M.~B.}\ \bibnamefont
  {d'Aug{\`e}res}}, \bibinfo {author} {\bibfnamefont {J.}~\bibnamefont
  {Zhang}}, \bibinfo {author} {\bibfnamefont {S.}~\bibnamefont {Janbaz}},
  \bibinfo {author} {\bibfnamefont {A.~M.}\ \bibnamefont {Arag{\`o}n}},\ and\
  \bibinfo {author} {\bibfnamefont {C.}~\bibnamefont {Coulais}},\ }\bibfield
  {title} {\bibinfo {title} {Fracture metamaterials with on-demand crack paths
  enabled by bending},\ }\href@noop {} {\bibfield  {journal} {\bibinfo
  {journal} {arXiv preprint}\ }\textbf {\bibinfo {volume} {2405.19061}}
  (\bibinfo {year} {2024})}\BibitemShut {NoStop}%
\bibitem [{\citenamefont {Gao}\ \emph {et~al.}(2024)\citenamefont {Gao},
  \citenamefont {Zhang}, \citenamefont {Wu}, \citenamefont {Pham},
  \citenamefont {Lu}, \citenamefont {Xia}, \citenamefont {Wang},\ and\
  \citenamefont {Wang}}]{gao2024damage}%
  \BibitemOpen
  \bibfield  {author} {\bibinfo {author} {\bibfnamefont {Z.}~\bibnamefont
  {Gao}}, \bibinfo {author} {\bibfnamefont {X.}~\bibnamefont {Zhang}}, \bibinfo
  {author} {\bibfnamefont {Y.}~\bibnamefont {Wu}}, \bibinfo {author}
  {\bibfnamefont {M.-S.}\ \bibnamefont {Pham}}, \bibinfo {author}
  {\bibfnamefont {Y.}~\bibnamefont {Lu}}, \bibinfo {author} {\bibfnamefont
  {C.}~\bibnamefont {Xia}}, \bibinfo {author} {\bibfnamefont {H.}~\bibnamefont
  {Wang}},\ and\ \bibinfo {author} {\bibfnamefont {H.}~\bibnamefont {Wang}},\
  }\bibfield  {title} {\bibinfo {title} {Damage-programmable design of
  metamaterials achieving crack-resisting mechanisms seen in nature},\
  }\href@noop {} {\bibfield  {journal} {\bibinfo  {journal} {Nature
  Communications}\ }\textbf {\bibinfo {volume} {15}},\ \bibinfo {pages} {7373}
  (\bibinfo {year} {2024})}\BibitemShut {NoStop}%
\bibitem [{\citenamefont {Karapiperis}\ and\ \citenamefont
  {Kochmann}(2023)}]{karapiperis2023prediction}%
  \BibitemOpen
  \bibfield  {author} {\bibinfo {author} {\bibfnamefont {K.}~\bibnamefont
  {Karapiperis}}\ and\ \bibinfo {author} {\bibfnamefont {D.~M.}\ \bibnamefont
  {Kochmann}},\ }\bibfield  {title} {\bibinfo {title} {Prediction and control
  of fracture paths in disordered architected materials using graph neural
  networks},\ }\href@noop {} {\bibfield  {journal} {\bibinfo  {journal}
  {Communications Engineering}\ }\textbf {\bibinfo {volume} {2}},\ \bibinfo
  {pages} {32} (\bibinfo {year} {2023})}\BibitemShut {NoStop}%
\bibitem [{\citenamefont {Kane}\ and\ \citenamefont
  {Lubensky}(2014)}]{kane2014topological}%
  \BibitemOpen
  \bibfield  {author} {\bibinfo {author} {\bibfnamefont {C.~L.}\ \bibnamefont
  {Kane}}\ and\ \bibinfo {author} {\bibfnamefont {T.~C.}\ \bibnamefont
  {Lubensky}},\ }\bibfield  {title} {\bibinfo {title} {Topological boundary
  modes in isostatic lattices},\ }\href@noop {} {\bibfield  {journal} {\bibinfo
   {journal} {Nature Physics}\ }\textbf {\bibinfo {volume} {10}},\ \bibinfo
  {pages} {39} (\bibinfo {year} {2014})}\BibitemShut {NoStop}%
\bibitem [{\citenamefont {Zhang}\ and\ \citenamefont
  {Mao}(2018)}]{zhang2018fracturing}%
  \BibitemOpen
  \bibfield  {author} {\bibinfo {author} {\bibfnamefont {L.}~\bibnamefont
  {Zhang}}\ and\ \bibinfo {author} {\bibfnamefont {X.}~\bibnamefont {Mao}},\
  }\bibfield  {title} {\bibinfo {title} {Fracturing of topological maxwell
  lattices},\ }\href@noop {} {\bibfield  {journal} {\bibinfo  {journal} {New
  Journal of Physics}\ }\textbf {\bibinfo {volume} {20}},\ \bibinfo {pages}
  {063034} (\bibinfo {year} {2018})}\BibitemShut {NoStop}%
\bibitem [{\citenamefont {Liu}\ \emph {et~al.}(2023)\citenamefont {Liu},
  \citenamefont {Sarkar}, \citenamefont {Arafat}, \citenamefont {Stanifer},
  \citenamefont {Gonella},\ and\ \citenamefont {Mao}}]{liu2023stress}%
  \BibitemOpen
  \bibfield  {author} {\bibinfo {author} {\bibfnamefont {H.}~\bibnamefont
  {Liu}}, \bibinfo {author} {\bibfnamefont {S.}~\bibnamefont {Sarkar}},
  \bibinfo {author} {\bibfnamefont {A.~N.}\ \bibnamefont {Arafat}}, \bibinfo
  {author} {\bibfnamefont {E.}~\bibnamefont {Stanifer}}, \bibinfo {author}
  {\bibfnamefont {S.}~\bibnamefont {Gonella}},\ and\ \bibinfo {author}
  {\bibfnamefont {X.}~\bibnamefont {Mao}},\ }\bibfield  {title} {\bibinfo
  {title} {Stress control in non-ideal topological maxwell lattices via
  geometry},\ }\href@noop {} {\bibfield  {journal} {\bibinfo  {journal} {arXiv
  preprint}\ }\textbf {\bibinfo {volume} {2311.18756}} (\bibinfo {year}
  {2023})}\BibitemShut {NoStop}%
\bibitem [{\citenamefont {Widstrand}\ \emph {et~al.}(2023)\citenamefont
  {Widstrand}, \citenamefont {Hu}, \citenamefont {Mao}, \citenamefont {Labuz},\
  and\ \citenamefont {Gonella}}]{widstrand2023stress}%
  \BibitemOpen
  \bibfield  {author} {\bibinfo {author} {\bibfnamefont {C.}~\bibnamefont
  {Widstrand}}, \bibinfo {author} {\bibfnamefont {C.}~\bibnamefont {Hu}},
  \bibinfo {author} {\bibfnamefont {X.}~\bibnamefont {Mao}}, \bibinfo {author}
  {\bibfnamefont {J.}~\bibnamefont {Labuz}},\ and\ \bibinfo {author}
  {\bibfnamefont {S.}~\bibnamefont {Gonella}},\ }\bibfield  {title} {\bibinfo
  {title} {Stress focusing and damage protection in topological maxwell
  metamaterials},\ }\href@noop {} {\bibfield  {journal} {\bibinfo  {journal}
  {International Journal of Solids and Structures}\ }\textbf {\bibinfo {volume}
  {274}},\ \bibinfo {pages} {112268} (\bibinfo {year} {2023})}\BibitemShut
  {NoStop}%
\bibitem [{\citenamefont {Widstrand}\ \emph {et~al.}(2024)\citenamefont
  {Widstrand}, \citenamefont {Mao},\ and\ \citenamefont
  {Gonella}}]{widstrand2024robustness}%
  \BibitemOpen
  \bibfield  {author} {\bibinfo {author} {\bibfnamefont {C.}~\bibnamefont
  {Widstrand}}, \bibinfo {author} {\bibfnamefont {X.}~\bibnamefont {Mao}},\
  and\ \bibinfo {author} {\bibfnamefont {S.}~\bibnamefont {Gonella}},\
  }\bibfield  {title} {\bibinfo {title} {Robustness of stress focusing in soft
  lattices under topology-switching deformation},\ }\href@noop {} {\bibfield
  {journal} {\bibinfo  {journal} {Extreme Mechanics Letters}\ }\textbf
  {\bibinfo {volume} {68}},\ \bibinfo {pages} {102135} (\bibinfo {year}
  {2024})}\BibitemShut {NoStop}%
\bibitem [{\citenamefont {de~Waal}\ \emph {et~al.}(2024)\citenamefont
  {de~Waal}, \citenamefont {Chouzouris},\ and\ \citenamefont
  {Dias}}]{dewaal2024architectingmechanismsdamagetopological}%
  \BibitemOpen
  \bibfield  {author} {\bibinfo {author} {\bibfnamefont {L.}~\bibnamefont
  {de~Waal}}, \bibinfo {author} {\bibfnamefont {M.}~\bibnamefont
  {Chouzouris}},\ and\ \bibinfo {author} {\bibfnamefont {M.~A.}\ \bibnamefont
  {Dias}},\ }\href {https://arxiv.org/abs/2410.17100} {\bibinfo {title}
  {Architecting mechanisms of damage in topological metamaterials}} (\bibinfo
  {year} {2024}),\ \Eprint {https://arxiv.org/abs/2410.17100} {arXiv:2410.17100
  [cond-mat.soft]} \BibitemShut {NoStop}%
\bibitem [{\citenamefont {Wang}\ \emph
  {et~al.}(2024{\natexlab{b}})\citenamefont {Wang}, \citenamefont {Sarkar},
  \citenamefont {Gonella},\ and\ \citenamefont
  {Mao}}]{Wang2024Topologicalmechanicalmetamaterial}%
  \BibitemOpen
  \bibfield  {author} {\bibinfo {author} {\bibfnamefont {X.}~\bibnamefont
  {Wang}}, \bibinfo {author} {\bibfnamefont {S.}~\bibnamefont {Sarkar}},
  \bibinfo {author} {\bibfnamefont {S.}~\bibnamefont {Gonella}},\ and\ \bibinfo
  {author} {\bibfnamefont {X.}~\bibnamefont {Mao}},\ }\href
  {https://doi.org/10.21203/rs.3.rs-5677605/v1} {\bibinfo {title} {Topological
  mechanical metamaterial for robust and ductile one-way fracturing}} (\bibinfo
  {year} {2024}{\natexlab{b}}),\ \Eprint {https://arxiv.org/abs/2410.17100}
  {Research Square:2410.17100} \BibitemShut {NoStop}%
\bibitem [{\citenamefont {Driscoll}\ \emph {et~al.}(2016)\citenamefont
  {Driscoll}, \citenamefont {Chen}, \citenamefont {Beuman}, \citenamefont
  {Ulrich}, \citenamefont {Nagel},\ and\ \citenamefont
  {Vitelli}}]{driscoll2016role}%
  \BibitemOpen
  \bibfield  {author} {\bibinfo {author} {\bibfnamefont {M.~M.}\ \bibnamefont
  {Driscoll}}, \bibinfo {author} {\bibfnamefont {B.~G.-g.}\ \bibnamefont
  {Chen}}, \bibinfo {author} {\bibfnamefont {T.~H.}\ \bibnamefont {Beuman}},
  \bibinfo {author} {\bibfnamefont {S.}~\bibnamefont {Ulrich}}, \bibinfo
  {author} {\bibfnamefont {S.~R.}\ \bibnamefont {Nagel}},\ and\ \bibinfo
  {author} {\bibfnamefont {V.}~\bibnamefont {Vitelli}},\ }\bibfield  {title}
  {\bibinfo {title} {The role of rigidity in controlling material failure},\
  }\href@noop {} {\bibfield  {journal} {\bibinfo  {journal} {Proceedings of the
  National Academy of Sciences}\ }\textbf {\bibinfo {volume} {113}},\ \bibinfo
  {pages} {10813} (\bibinfo {year} {2016})}\BibitemShut {NoStop}%
\bibitem [{\citenamefont {Mao}\ and\ \citenamefont
  {Lubensky}(2018)}]{mao2018maxwell}%
  \BibitemOpen
  \bibfield  {author} {\bibinfo {author} {\bibfnamefont {X.}~\bibnamefont
  {Mao}}\ and\ \bibinfo {author} {\bibfnamefont {T.~C.}\ \bibnamefont
  {Lubensky}},\ }\bibfield  {title} {\bibinfo {title} {Maxwell lattices and
  topological mechanics},\ }\href@noop {} {\bibfield  {journal} {\bibinfo
  {journal} {Annual Review of Condensed Matter Physics}\ }\textbf {\bibinfo
  {volume} {9}},\ \bibinfo {pages} {413} (\bibinfo {year} {2018})}\BibitemShut
  {NoStop}%
\bibitem [{\citenamefont {Lubensky}\ \emph {et~al.}(2015)\citenamefont
  {Lubensky}, \citenamefont {Kane}, \citenamefont {Mao}, \citenamefont
  {Souslov},\ and\ \citenamefont {Sun}}]{lubensky2015phonons}%
  \BibitemOpen
  \bibfield  {author} {\bibinfo {author} {\bibfnamefont {T.}~\bibnamefont
  {Lubensky}}, \bibinfo {author} {\bibfnamefont {C.}~\bibnamefont {Kane}},
  \bibinfo {author} {\bibfnamefont {X.}~\bibnamefont {Mao}}, \bibinfo {author}
  {\bibfnamefont {A.}~\bibnamefont {Souslov}},\ and\ \bibinfo {author}
  {\bibfnamefont {K.}~\bibnamefont {Sun}},\ }\bibfield  {title} {\bibinfo
  {title} {Phonons and elasticity in critically coordinated lattices},\
  }\href@noop {} {\bibfield  {journal} {\bibinfo  {journal} {Reports on
  Progress in Physics}\ }\textbf {\bibinfo {volume} {78}},\ \bibinfo {pages}
  {073901} (\bibinfo {year} {2015})}\BibitemShut {NoStop}%
\end{thebibliography}
%

\clearpage

\appendix

\section{Supplementary Information}

In this supplementary information, we provide additional results for the experiments conducted on the lattices. Additionally, we present a numerical imperfection study for the RDP1-LUP1 lattice to elucidate the influence of material imperfections on the damage process. 

\begin{figure*}
\includegraphics{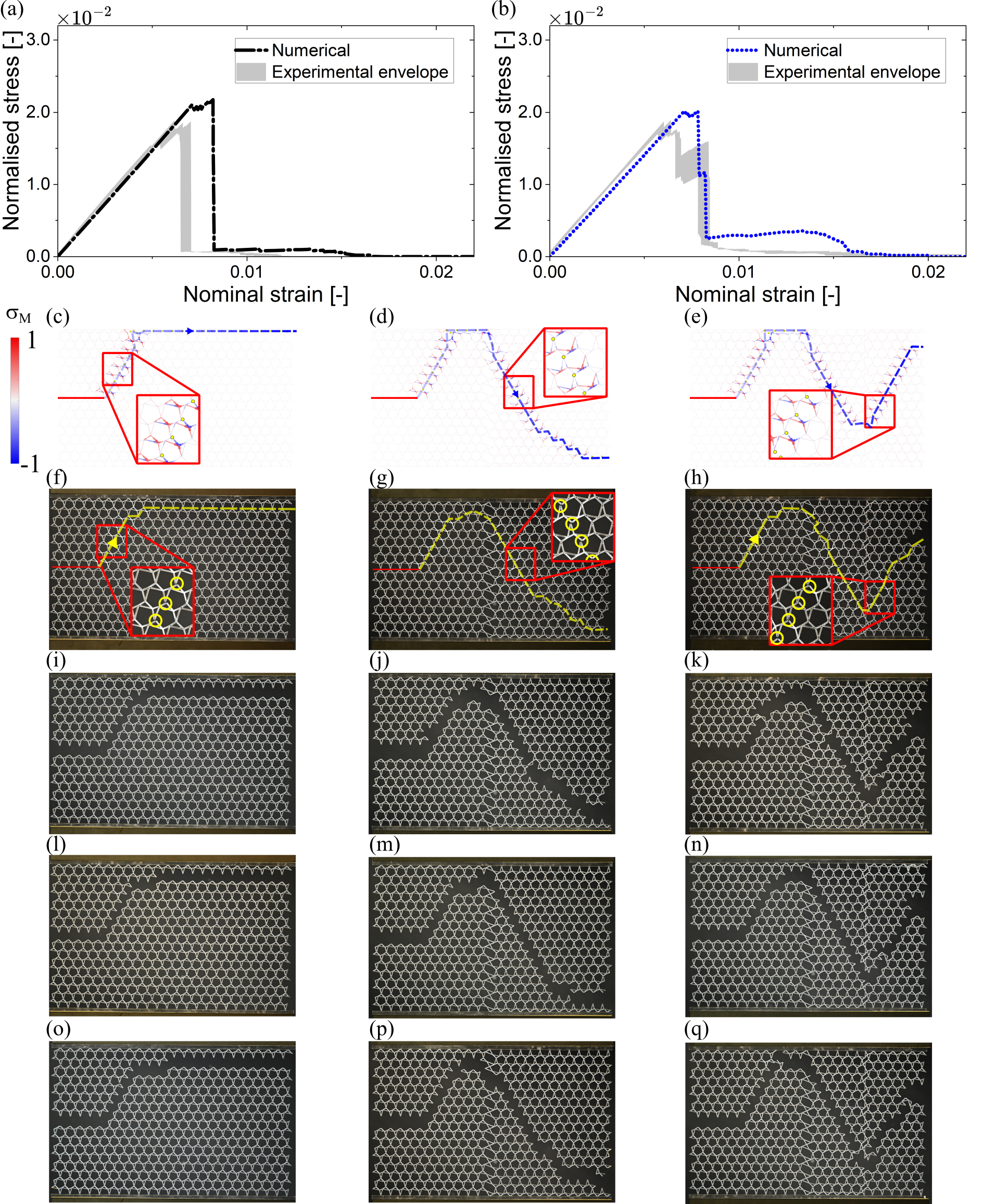}
\caption{\label{fig:S1} Numerical vs experimental results. Normalised stress vs nominal strain for (a) RDP1; and, (b) RDP1-LUP1-RDP1. Initial damage path with bridging elements intact for (c,f) RDP1; (d,g) RDP1-LUP1; and, (e,h) RDP1-LUP1-RDP1. Complete fracture observed in experiments for: (i,l,o) RDP1; (j,m,p) RDP1-LUP1; and, (k,n,q) RDP1-LUP1-RDP1. In (c-e), $\sigma_\mathrm{M} = \sigma_\mathrm{A}/\sigma_\mathrm{A}^\mathrm{u}+\mathrm{sgn}(\sigma_\mathrm{A})\left|\sigma_\mathrm{B}/\sigma_\mathrm{B}^\mathrm{u}\right|$ plots the maximum normalised stress within each element and yellow circles represent locations where the node has split.}
\end{figure*}

\textbf{Experimental results - } This section summarises the experiments not reported in the main text. All experiments were carried out at a 0.5mm/min loading rate with fixed boundary constraints. Polarised light was placed behind the lattice and a polarised lens was placed on the camera lens to indicate the force chains that develop within the lattice. Fig. \ref{fig:S1}(a,b) provides the normalised stress vs strain comparison for RDP1 and RDP1-LUP1-RDP1. As highlighted in the main text, whilst the peak stress is lower for the experiments, the general behaviour observed matches the numerical prediction well. In the initial stretching dominated phase, SSS are removed as the damage propagates in the direction of the FM that localise at the crack-tip (Fig. \ref{fig:S1}(c,g)). As the crack propagates, bridging elements remain behind the crack path. These elements are responsible for the energy absorbed during the bending dominated phase. These sequentially fail, leading to the complete separation of the lattice (Fig. \ref{fig:S1}(d-f,h-j)). 

\begin{figure}
\includegraphics{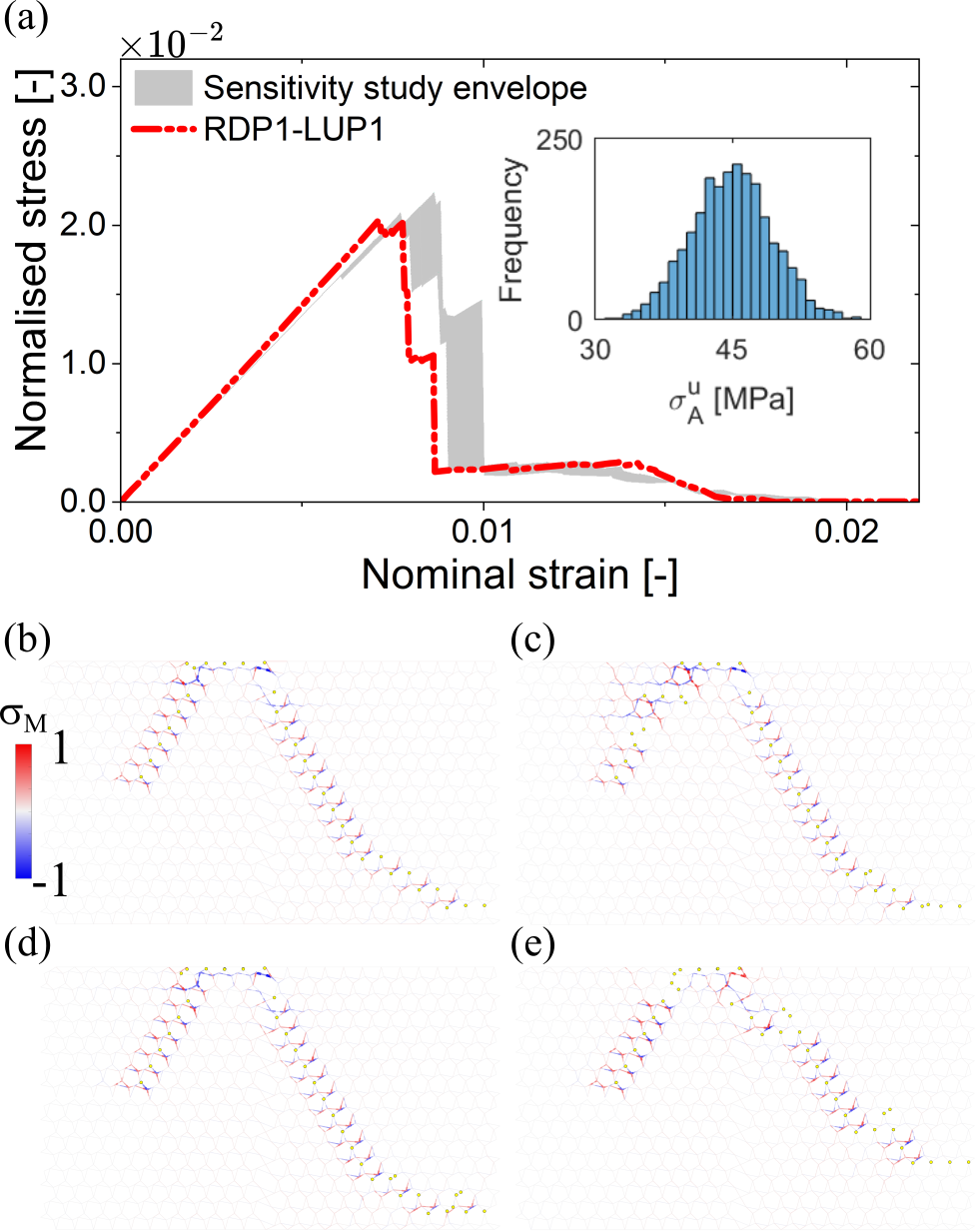}
\caption{\label{fig:s2} Numerical sensitivity study for RDP1-LUP1. (a) Normalised stress vs nominal strain and fracture energy comparison. The small insert provides an example of one of the simulation's ultimate axial bond stress distribution ($\sigma_\mathrm{A}^\mathrm{u}$) with a mean of 45MPa and a standard deviation of 4.5MPa. Lattice following initial damage propagation for: (a) no imperfections; and, (c-e) imperfections.}
\end{figure}

\textbf{Imperfection study - } A numerical sensitivity study was conducted applying a normal distribution to $\sigma_\mathrm{A}^\mathrm{u}$ (refer to the insert in Fig. \ref{fig:s2}(a)) and $\sigma_\mathrm{B}^\mathrm{u}$ of the bonds (\emph{i.e.} the same ultimate stress value is set for the three elements that make up a bond). The standard deviation was set to 10\% of the prescribed values. The results from three simulations are compared with the simulation with no imperfections in Fig. \ref{fig:s2}. In these simulations, the general response remained similar, however, an increase in the fracture energy of the lattice was observed when these material imperfections were included (Fig. \ref{fig:s2}(a)). 

In all cases the damage path followed a similar trend, propagating diagonally upwards from the crack tip to the boundary and then diagonally downwards after passing through the domain wall. Highlighting the damage process is robust against material imperfections within the system.

\end{document}